\documentclass[conference]{IEEEtran}
\usepackage{graphics}
\usepackage{graphicx}
\usepackage{epsfig,amssymb}
\usepackage{epstopdf}
\usepackage{amsmath}
\usepackage{breqn}
\usepackage{times}
\usepackage{mathrsfs}
\usepackage{array}
\usepackage{amsmath, latexsym, amsfonts, amssymb}
\usepackage[mathscr]{eucal}
\usepackage{array, tabularx}
\usepackage[table]{xcolor}
\usepackage{multirow}
\usepackage{epsfig}

\usepackage{cite}
\usepackage[numbers,sort & compress]{natbib}

\usepackage{tikz}
\usepackage{url}
\usepackage{mathtools}
\usepackage{psfragx}
\usepackage{xcolor}
\usepackage{bbm}
\usepackage{authblk}
\usepackage{balance}

\newcommand{\paren}[1]{\left(#1\right)}
\newcommand{\sqparen}[1]{\left[#1\right]}
\newcommand{\brparen}[1]{\left\{#1\right\}}

\newcommand{\R}{\ensuremath{\field{R}}} % real numbers
\newcommand{\C}{\ensuremath{\field{C}}} % real numbers
\newcommand{\defeq}{\ensuremath{\triangleq}} %Triangle equation for definitions
\renewcommand{\vec}[1]{\ensuremath{\boldsymbol{#1}}} %Re-define \vec command to generate vectors in bold
\newcommand{\ie}{\ensuremath{{\text{\em i.e.}}}}

\newcommand{\D}{\displaystyle}

\newtheorem{theorem}{Theorem}
\newtheorem{lemma}{Lemma}

\IEEEoverridecommandlockouts

\def\B{
	\begin{bmatrix}
		1 \\
	e^{j\theta_1} \\
		0
	\end{bmatrix}}
	
\def\C{
	\begin{bmatrix}
		1 \\
	e^{j\theta_2} \\
	0
	\end{bmatrix}}
		
\def\D{
	\begin{bmatrix}
		1 \\
	e^{j\theta_3} \\
		0
	\end{bmatrix}}
	
\def\E{
	\begin{bmatrix}
		1 \\
		0 \\
		0
	\end{bmatrix}}

	\def\P{
	\begin{bmatrix}
		1 \\
	0\\
		e^{j\theta_1} 
	\end{bmatrix}}
	
\def\Q{
	\begin{bmatrix}
		1 \\
		0\\
		e^{j\theta_2} 
\end{bmatrix}}
		
\def\R{
	\begin{bmatrix}
		1 \\
		0\\
		e^{j\theta_3} 
	\end{bmatrix}}
	
\def\S{
	\begin{bmatrix}
		1 \\
	0 \\
		0
	\end{bmatrix}}		
	
\begin{document}

%\title{Cluster-based Opportunistic  Wireless Energy Beamforming  using RSSI Feedback}
%\title{Exploiting Multiuser Diversity with RSSI Feedback for Wireless Power Transfer}
\title{Cluster-based Wireless Energy  Transfer for Low Complex Energy Receivers}

\author[$\ast \dag $]{Samith Abeywickrama}
\author[$ \ddag $]{Tharaka Samarasinghe}
\author[$ \dag $]{Chau Yuen}
\author[$ \ast  $]{Rui Zhang}
\affil[$ \dag $]{Singapore University of Technology and Design,  Singapore}
\affil[$ \ddag $]{University of Moratuwa, Sri Lanka}
\affil[$ \ast $]{National University of Singapore, Singapore 
	
	\thanks{This work was supported in part by A*Star SERC project 1420200043, and in part by the National Natural Science Foundation of China (NSFC) 61750110529.}

}

\affil[$  $ ]{Email: \textit { tharindu@mymail.sutd.edu.sg, tharaka@ent.mrt.ac.lk, yuenchau@sutd.edu.sg, elezhang@nus.edu.sg  } }

\date{}
\bibliographystyle{ieeetr}
\maketitle

\begin{abstract}
	
This paper proposes a novel channel estimation method and a cluster-based opportunistic scheduling
policy, for a wireless energy transfer (WET)  system consisting of multiple low-complex energy receivers (ERs) with limited processing capabilities.	
Firstly, in the training stage, the energy transmitter (ET) obtains a set of Received Signal
Strength Indicator (RSSI) feedback values from all ERs, and these values are used to estimate the channels between the ET and all ERs.
Next, based on the channel estimates, the ERs are grouped into clusters, and the cluster that
has its members closest to its centroid in phase is selected for dedicated WET.  The beamformer that maximizes the minimum harvested energy among all ERs in the selected cluster is found by solving a convex optimization problem.  All ERs have the same chance of being selected regardless of their distances from the ET, and hence, this scheduling policy can be considered to be opportunistic as well as fair. It is shown that the proposed method achieves significant performance gains over benchmark schemes.
\end{abstract}

\section{Introduction}

Radio frequency (RF) signal enabled wireless energy transfer (WET) using multiple antennas at the energy transmitter (ET) has become a promising technology for enabling  convenient and perpetual power supply to  freely located wireless devices \cite{power1}. Increasing the efficiency of the energy transfer between the ET and the energy receiver (ER) is of paramount importance in WET. When multiple ERs are present, the availability of channel state information (CSI) and the scheduling policy have great impact on the achievable  efficiency. To this end, this paper has two main contributions. Firstly, it proposes a novel and practical  channel estimation method based on Received Signal
Strength Indicator (RSSI) feedback values from ERs, which allows the ET to perform optimal beamforming for the energy transfer. Secondly, it introduces a new cluster-based opportunistic scheduling policy, which enhances the efficiency of a WET system consisting of multiple ERs, while ensuring fairness.

WET based on the  CSI at the ET practically requires  a training stage for channel learning. However, due to tight energy constraints and hardware limitations in most practically available ERs, the conventional pilot-based techniques, where the channel estimation or the signal-to-interference-plus-noise ratio  (SINR)  calculation is done at the ERs, give rise  to many implementation difficulties. These difficulties thus call for new and low-complex channel learning methods, which  are deemed to be particularly useful  for WET.

To this end, the authors of \cite{one_bit} propose  to estimate the channel at the ET using a one-bit feedback algorithm, where phase perturbations are made based on the feedback bits to obtain a satisfactory beamforming vector for WET. In \cite{bruno1},  a novel waveform design strategy is proposed  by relaxing the assumption of perfect CSI at the ET. 
\cite{recip3} proposes exploiting the channel reciprocity for channel learning. That is, the ET determines the CSI of the forward link by estimating that  of the reverse link between the ER and the ET. Being different to our work, this method is mainly applicable for time division duplex (TDD) systems that
use the same frequency for the uplink and the downlink. Also, using channel reciprocity for channel estimation leads to many practical difficulties, due to the non-symmetric characteristics of the RF front-end circuitry at the receiver and the transmitter \cite{samith_noma}. Prior works  \cite{rssi_work, our_rssi, our_tsp} propose energy efficient channel estimation methods based on RSSI values that are fed back from the receiver to the transmitter, and among  them, \cite{our_rssi,our_tsp} can be considered to be the most related to our work. Specifically, \cite{our_rssi,our_tsp}  propose an approach of estimating the phase values of the channels between a single ER and each antenna of the ET, and these estimates are utilized to employ equal gain transmit (EGT) beamforming for WET. In this paper, we focus on utilizing RSSI values to estimate both channel phase and channel magnitude information using a maximum likelihood analysis, in order to perform more superior maximum ratio transmit (MRT) beamforming for the energy transfer in  single-ER case. 

Since we consider multiple ERs (unlike single ER in  \cite{our_rssi,our_tsp}) in general, an important  scheduling problem for WET needs to be solved. To this end, \cite{oppor1}  proposes an opportunistic scheduling policy for WET, where the beamformer is designed based on the ER having the best channel. This method increases the amount of energy transferred compared to the conventional round-robin scheduling. In \cite{oppor2}, random beamforming,  where the ET randomly selects a beamformer regardless of the channel information of the ERs, is proposed. This random selection ensures fairness. In \cite{user_clustered}, a novel user-clustered opportunistic beamforming  \cite{Tse02, hassibi, 6857357, 7404030} scheme is employed by utilizing the SINR values that are obtained from the receivers. \cite{oppor1,oppor2,user_clustered} are significantly different to our work as the low-complex ERs in our setup can only feed back  RSSI values to the ET.  Also, our proposed cluster based scheduling policy is both opportunistic and fair.

The clustering algorithm in this paper stems on the idea presented in \cite{chanaka_wiopt}, where the optimality of WET to a pair of low complex ERs is studied. In particular, \cite{chanaka_wiopt} highlights that the WET will be more efficient when the two ERs are close to each other in terms of channel phase. Along these lines, we  group the ERs into clusters using the Lloyd's Algorithm \cite{lloyd} by utilizing the channel phase estimates between the ET and the ERs. We pick the cluster that has its members (ERs) closest to its centroid in phase. Although all ERs in the network will harvest energy when performing the WET, we  give priority to this selected cluster, as the insights from \cite{chanaka_wiopt} suggest that the system will do better in terms of WET by focusing on this cluster compared to a random selection of ERs.   This makes our scheduling policy  opportunistic. The scheduling policy also ensures that all ERs in the network have the same chance of being in the selected cluster, regardless of their distances from the ET, thus making it fair over time as well. After cluster selection,  we solve a convex optimization problem to find the beamformer that maximizes the minimum harvested energy among the cluster members. This additionally  ensures intra-cluster fairness.  

The paper is organized  as follows. The system model and the problem formulation are presented in Section \ref{Section:System model}. Section \ref{tra} discusses the channel estimation, and Section \ref{opt} discusses how the optimization problem can be solved by utilizing the estimates. Then, in Section \ref{results}, we demonstrate the significant performance improvements that can be obtained thanks to the proposed estimation method and the scheduling policy, through simulations. Section \ref{conclusions} concludes the paper.

\section{System Model and Problem Setup}\label{Section:System model}

We consider a multiple-input single output (MISO)  channel for WET. An ET consisting of $K \geq 2$ antennas delivers energy over a wireless medium to $N$ ERs, each equipped with a single antenna. %The cell is modeled as a disk with the ET located at the center of the
%disk. We assume the ERs to be located uniformly over the plane of the cell.  %, see Fig. \ref{sm}.
%We assume $N >> K$. 
The ET in general transmits  $ M\leq K $ beams along the direction of $M$ beamforming vectors $\brparen{\vec{b_m} \in \mathbb{C}^{K \times 1}}^M_{m=1}$, such that the transmit signal at the ET is given by $$ \vec{x} = \sum_{m=1}^M\vec{b}_m s_m, $$ where $ s_m $ denotes the transmit symbol, which is independent over $ m $, and $ \mathbb{E}( |s_m|^{2}) =1$, $ \forall m  $.  
%= \mathbb E  (\vec{w}\vec{w}^{\dag})$, where $ \dag $ denotes the conjugate transpose. $ \mathrm {\bf C}_{\vec {xx}} $ is positive semi-definite, thus the number of energy beams $ d $ can be obtained from the 
%rank of $ \mathrm {\bf C}_{\vec {xx}} $ \cite{power1}, \textit{i.e.,} $ d= \mathrm{rank}(\mathrm {\bf C}_{\vec {xx}}) $ . 
It is assumed that the maximum transmit sum-power constraint at the ET is $ P > 0 $.  Therefore, we have $ \mathbb{E} (\|\vec{x}\|^{2}) = \mathrm{tr}(\mathrm {\bf C}_{\vec {xx}}) \leq P$, where $  \mathrm {\bf C}_{\vec {xx}} = \mathbb E (\vec{x}\vec{x}^{\dag})$ is the transmit covariance  matrix, and $\mathrm{tr}(\cdot)$ and $ \|\cdot \| $ denote the trace of a square matrix and the Euclidean norm, respectively.

Let $ \vec{h}_i  \in \mathbb{C}^{K \times 1}$ represent the random complex MISO channel vector between the ET and the $i$-th ER, such that $\vec{h}_i = \sqparen {|h_{i,1}|e^{j\delta_{i,1}}, \dots , |h_{i,K}|e^{j\delta_{i,K}}}^\top $. For the simplicity
in notations, $\vec{h}_i$ is assumed to be the product of the path
loss and multipath fading between the ET and the $i$-th ER. The channel magnitudes are considered to be independent and identically distributed
(i.i.d.) with an arbitrary distribution, and the channel phase values are considered to be uniformly distributed between 0 and 2$\pi$. The received energy (or RSSI) at the $i$-th ER can be written as
\begin{eqnarray}
	\mathrm R_i = \xi (\vec h^{\dag}_i \mathrm {\bf C}_{\vec {xx}}\vec h_i), \label{power}
\end{eqnarray}
where $\xi$ denotes the conversion efficiency of the ER \cite{power1}. We assume $ \xi=1 $ for  simplicity, %and $z_i$ represents the effect of noise on $  \mathrm R_i $.
%Note that due to noise, the RSSI value will change from one measurement to the other. We use random variable $ z_{i}$ to represent the effect of noise on $  \mathrm R_{i} $.  More specifically,  $ z_{i} $ captures the effect of  all noise related to the measurement process such as noise in the channel, circuit, antenna matching network and rectifier. 
and consider a quasi-static block-fading channel model and a block-based energy transmission, where it is assumed that $\vec{h}_i$ remains constant over each transmission block. %More specifically, we assume that the channel is slowly varying so that during each transmission block, $ \vec{h}_i $ can be considered to be unknown, but non varying (fixed). 

%Therefore, we assume that in a given transmission block, the randomness in \eqref{power} is caused only by the noise component $z_{i}$, which we assume to be additive Gaussian with zero mean and variance $\sigma^2$. %The transmission block has a length $ T > 0 $. 

%amplitude multipath factor is the multipath induced phase

It is well known that CSI plays a vital role in beamforming. Therefore, the WET process consists of two stages. Firstly, we have the training stage that the ET uses for channel learning. Then, the knowledge on the channel is used to set the beamforming vectors for the second stage, that we call the wireless power beamforming (WPB) stage. This is where the actual WET is conducted. 
%For the $i$-th ER, this can be done if the ET has the knowledge of $ \vec{h}_i $.
%In practice, each transmit antenna has its own power amplifier which operates properly only when the transmit power is below a pre-designed threshold. Therefore, we employ equal gain transmit (EGT) beamforming in the WPB stage, where the ET equally splits the power among all transmit antennas and pre-compensates channel phase shifts such that the signals are coherently added up at the ER \cite{egc2}. This means, the ET designs the beamformer by assuming that the magnitudes of the $K$ channels between the ET and a selected ER are equal, and it allocates equal transmit power to all $ K $ antennas such that the maximum transmit sum-power constraint is satisfied. 
%For the $i$-th ER, this can be done if the ET has the knowledge of $\brparen{|h_{i,1}|}_{k=2}^K$ and $\brparen{\phi_{i,k}}_{k=2}^K$, where $ \phi_{i,k} = \delta_{i,k} - \delta_{i,1} $. $\phi_{i,k}$ represents the phase difference of the two channels between the ER and the first antenna of the ET, and the $k$-th antenna of the ET. Here, the first antenna is selected as reference without any loss of generality. 
%In practice, this can be achieved by estimating the channel at the ER, and feeding back the channel information to the ET. However, 
Since we  particularly focus on applications having tight energy constraints at the ERs, performing channel estimation at the ER directly may become infeasible, as it involves analog to digital conversion and baseband  processing, which require significant energy.  Therefore, we focus on obtaining estimates of $\brparen{\vec{h}_i}_{i=1}^{N}$ by only considering RSSI values that are fed back from the ERs to the ET. In most receivers, the RSSI values are in fact already available, and no significant signal processing is needed to obtain them. The utility of these estimates are mainly twofold. Firstly, we use these channel estimates to group the ERs into clusters. Then, we use them to  perform multi-user optimal beamforming in the WPB stage.

%\begin{figure}[t] 
%    \centering {\includegraphics[scale=1]{sm4}} 
%    \caption{System model.}     
%    \label{sm}
%\end{figure}

%Let $ \hat {\vec{h}}_i  $ denotes the estimated channel vector from the training stage for all $i \in \brparen{1,\ldots,N}$. 
%As discussed earlier, only the channel phase differences are required to estimated for EGT beamforming. The channel magnitudes of the $K$ channels between the ET and a selected ER is irrelevant for the beamfomokming process, and hence, not estimated. Therefore, we assume that $ \hat {\vec h}_i $ takes the form of $ \sqparen {1, \hat \phi_{i,2}, \dots , \hat \phi_{i,K}}^\top $. Since there are multiple ERs with $N >>K$, we will use these estimates to define $ M $ clusters of ERs through vector quantization, and each ER will be represented by the centroid of its cluster. 

The ER clustering is as follows. We cluster the ERs into $Q$ clusters, based on the {\em phase values of the estimated channel vectors} (to be specified in Section III). For $i \in \brparen{1, \ldots, N}$, $\vec{\delta}_i = \sqparen {{\delta_{i,1}}, \dots , {\delta_{i,K}}}^\top $, \textit{i.e.}, the vector containing all phase values in $\vec{h}_i$.  We  partition $\brparen{\vec{\delta}_i}_{i=1}^N$ into $Q$ clusters, denoted by $ \mathrm{\mathbf S} = \{S_1, \dots , S_Q \} $, by minimizing the intra-cluster sum of squares,  given by,
\begin{equation}
\underset{  \footnotesize{  \textrm {\textbf{S}} }   }{ \text{arg min}}  \sum_{q=1}^{{Q}} \sum_{\vec \delta_i \in S_q} \| \vec \delta_i - \vec w_q \|^2, \label{algo}
\end{equation}
where $$ \vec w_q = \frac{1}{ |S_q|} \sum_{\vec \delta_i \in S_q} \vec \delta_i $$ denotes the  centroid of $q$-th cluster. This is also equivalent to maximizing the squared deviations between members of different clusters as well \cite{Kriegel2017}. This is an NP-hard problem, and we use Lloyd's algorithm \cite{lloyd} to obtain the solution. It should be noted that ERs in the same cluster may not be close to each other spatially since the clustering is done based on the phase values of the channel vectors. Let ${S}^\star$ be the cluster that has its cluster members (ERs) closest to its centroid, \textit{i.e.},
\begin{equation}
S^\star= \underset{  \footnotesize{ S_q \in \textrm {\textbf{S}}  }   }{ \text{arg min}}   \sum_{\vec \delta_i \in S_q} \| \vec \delta_i - \vec w_q \|^2.
\end{equation}
All the ERs in the network will harvest energy in the WPB stage. However, we will give priority to the ERs in ${S}^\star$, as the insights from \cite{chanaka_wiopt} suggest that the system will do better in terms of WET by focusing on this cluster compared to a random selection of ERs.  
This makes the algorithm opportunistic. Moreover, we formulate an optimization problem to design a beamformer that maximizes the minimum harvested energy among all ERs in ${S}^\star$, with a goal of being fair among the cluster members  in ${S}^\star$ as well.

The clustering is done by only considering the phase values of the estimated channel vectors due to the following  reasons. Firstly, since the phase values change rapidly over time (i.i.d. in our model), all ERs have the same chance of being in the selected cluster, which ensures fairness for the whole network over time. If the magnitudes of $\brparen{\vec{h}_i}_{i=1}^{N}$ are considered, the location dependent path loss values of the ERs, which  change slowly over time, will play a significant role  in  clustering, and thus, will affect the fairness in scheduling. Secondly, due to the phase values being uniformly and identically distributed, the cluster sizes will not differ significantly from each other. Note that, the sum of Euclidean distances between the ERs and a centroid of the cluster depends on the number of ERs in the cluster. Therefore, if there is a large variation in cluster sizes, ${S}^\star$ may end up being the smallest cluster with the lowest number of ERs, and this will not serve our purpose as well. We should stress that the notion of fairness in this paper is providing each ER in the network equal opportunity for being in $S^\star$ regardless of its distance from the ET, and being prioritized in the WPB stage. The harvested energy will differ among ERs depending on their distances from the ET.

Let $ \hat {\vec{h}}_i  $ denote the estimated channel vector of ER $i \in {S}^\star$, and let $ \vec \eta_i $ denote the channel estimation error.  $ \vec \eta_i $ is assumed to be bounded, \textit{i.e.}, $\|\vec \eta_i \|_{\mathrm F} = \sqrt{\vec \eta_i^{\dag} \vec \eta_i }  \leq \varepsilon_i ,$
%\begin{equation}
% \|\vec \eta_i \|_{\mathrm F} = \sqrt{\vec \eta_i^{\dag} \vec \eta_i }  \leq \varepsilon_i ,
%\end{equation}
where $  \| \cdot \|_{\mathrm F} $ denotes the Frobenius norm and $ \varepsilon_i \geq 0 $.\footnote{ It should be noted that according to the estimation methodology in this paper (discussed in Section III), the estimation error is in fact unbounded, and a probabilistic constraint may have  been more suitable. We have assumed bounded channel estimation uncertainties for the analytical tractability of the problem.  Please refer to \cite{bounded} where a similar approximation is made, and the necessity and the fairness of the approximation are justified.} By using these notations, the received energy (or RSSI) at the $i$-th ER in ${S}^\star$ can be written as $  (\hat {\vec h}_i + \vec \eta_i)^{\dag} \mathrm {\bf C}_{\vec {xx}} (\hat {\vec h}_i + \vec \eta_i) $. Thus, our optimization problem can be formulated as 
%\footnote{ It should be noted that the estimation error is actually unbounded since we have considered Gaussian noise, and a probabilistic constraint may have  been more suitable. We have assumed bounded channel estimation uncertainties for the analytical tractability of the problem.  Please refer to \cite{bounded} where a similar approximation is made, and the necessity and the fairness of the approximation is justified.} 
%Now, considering the $m$-th cluster centroid,  we write an expression for the received energy (or RSSI) assuming unit channel magnitudes, \textit{i.e.}, $ \xi (\hat {\vec h}_m + \vec \eta_m)^{\dag} \mathrm {\bf C}_{\vec {xx}} (\hat {\vec h}_m + \vec \eta_m) $. Note that since EGT beamforming is used, this assumption does not effect the solution of the optimization problem, which can be formulated as
\begin{equation}
	\begin{aligned}
		& \underset{\footnotesize \mathrm {\bf C}_{\vec {xx}} \succeq 0,\ t \geq 0  }{\text{maximize}}
		& & t  \\
		& \text{subject to}
		& & \hspace{-0.25cm}\mathrm C1: \hspace{-0.25cm} \underset{\footnotesize \|\vec \eta_i \|  \leq \varepsilon_i }{\text{min}}  (\hat {\vec h}_i + \vec \eta_i)^{\dag} \mathrm {\bf C}_{\vec {xx}} (\hat {\vec h}_i + \vec \eta_i) \geq t \ \ \forall i \in {S}^\star \\
		& &&  \hspace{-0.25cm} \mathrm C2: \mathrm{tr}(\mathrm {\bf C}_{\vec {xx}}) \leq P, 
	\end{aligned}\label{Optimization Problem-MU} 
\end{equation} 
where $ t $ is a real-valued optimization variable. The problem is convex, but it is complex due to $ \mathrm C1 $ having infinitely many inequalities. Also, it has been shown in \cite{rank_one}, that if an optimization problem of the form in \eqref{Optimization Problem-MU} is solvable, the rank of the solution is one, $\ie$, optimality is achieved when $\mathrm{rank}(\mathrm {\bf C}_{\vec {xx}})=1$. This means that, it is optimal to transmit a single beam in the downlink for WET. This sheds further light into why clustering will be useful in this context, as we will be better off focusing on a set of ERs that are closer to each other in terms of their channel phase values, than considering all ERs, when setting  the beamforming vector.  

In the next section, we  discuss how the estimates of $\brparen{ {\vec{h}}_i}_{i=1}^{N}$ can be obtained by using RSSI feedback values, and how the estimates can be utilized for clustering.  Then, in Section \ref{opt} we solve the optimization problem of interest.

\section{Training Stage and channel estimation} \label{tra}

In \cite{our_rssi}, $K$ and $N$ are assumed to be 2 and 1, respectively, and a method of utilizing RSSI feedback values to estimate the phase difference of the two MISO channels between the ET antennas  and the ER antenna has been proposed. This method can be directly extended to estimate $\brparen{\phi_{i,v}}_{v=2}^K$, where $ \phi_{i,v} = \delta_{i,v} - \delta_{i,1} $. That is, for a given ER $i$, we can estimate all phase values of the channel vector $\vec{h}_i$, relative to the phase value of the channel between the ER and the first antenna of the ET (the first antenna is selected as reference without any loss of generality). Therefore, since only phase information is available, the training and estimation schemes  proposed in \cite{our_rssi}, and \cite{our_tsp} (where an extension for $K>2$ is proposed), can only be used to employ EGT beamforming in the WPB stage. With EGT beamforming, the ET equally splits the power among the transmit antennas, and pre-compensates channel phase shifts such that the signals are coherently added up at the ER, regardless of the channel magnitudes.
Since we are interested in employing optimal beamforming in the WPB stage, the ET has to estimate $\brparen{\phi_{i,v}}_{v=2}^K$ as well as the channel magnitudes $\brparen{|h_{i,k}|}_{k=1}^K$ of all $i \in \brparen{1, \ldots, N}$. Hence, we need modified training and estimation schemes to facilitate these improvements.

\begin{figure}[t] \vspace{0.2cm}
	\centering {\includegraphics[trim = 0mm 0mm 0mm 0mm, clip,scale=1]{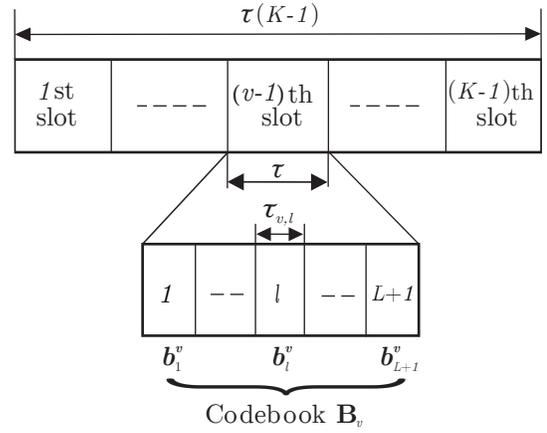}} 
	\caption{Training Stage.}	 
	\label{2pha} %\vspace{-0.5cm}
\end{figure} 

We  start by defining a set of codebooks. For $v \in \brparen{2, \ldots, K}$, we define codebook $ \mathrm {\bf {B}}_v = \sqparen{\vec{b}_{1}^v \ 	\ldots \ \vec{b}_{L+1}^v} $ that includes $(L+1) $ complex $K$-by-$1$ beamforming vectors. To this end, the $l$-th element of $\mathrm {\bf {B}}_v $ takes the form of
\begin{equation*}
\vec{b}_{l}^v = \sqrt{\frac{P}{2}} \sqparen {b_{l,1}^v\ \   \cdots\ \  b_{l,K}^v}^\top .
\end{equation*} 
Moreover, we have
\begin{alignat}{1}
b_{l,k}^v = \begin{cases}
1   & \text{if } k=1  \\
\exp\paren{j\theta_l}    & \text{if } k=v \ \ \text{and} \ \ l \neq L+1  \\
0  & \text{otherwise}  \\
\end{cases},
\end{alignat}
where $ \theta_l =  \frac{2(l-1)\pi}{L} $. 

The training stage consists of $K-1$ time slots as illustrated in Fig. \ref{2pha}. Each time slot is further subdivided into $L+1$ minislots. In the $L+1$ minislots in time slot $v-1$, the ET sequentially transmits using the $L+1$ beamforming vectors in $\mathrm {\bf {B}}_v $, and repeats for all $v \in \brparen{2, \ldots, K}$. Each ER will measure the respective $(K-1) \times (L+1)$ RSSI values for this sequential transmission, and  feed them back to the ET over orthogonal feedback channels \cite{orthoganal_feedback}. Note that the time taken for the channel learning does not depend on $ N $, and it depends only on $ K $ and $L$. 

For  clarity, we will explain the structure of the codebooks by using an example. Consider that $ K = 3 $ and $ L=3 $. For this selection, we have
 
\begin{equation*}
\mathrm {\bf {B}}_2  = \sqrt{\frac{P}{2}} \left(\B \C \D \E \right)_{3 \times 4}, 
\end{equation*}
and
\begin{equation*}
\mathrm {\bf {B}}_3  = \sqrt{\frac{P}{2}} \left(\P  \Q \R \S \right)_{3 \times 4}. 
\end{equation*}
There are two time slots in the training stage, and there are four minislots in each time slot. The ET will sequentially transmit using the eight beamforming vectors in $\mathrm {\bf {B}}_2$ and $\mathrm {\bf {B}}_3$ in the eight minislots. From this example,
it is not hard to see that for all beamforming vectors except the last one in each codebook, the ET employs a pairwise antenna activation policy. To be more general, for the first $L$ beamforming vectors transmitted in time slot $v-1$, the ET only activates the first antenna and the $v$-th transmit antenna, for all $v \in \brparen{2, \ldots, K}$.  
%In the training stage, the ET transmits only one beam simultaneously. That is, all the transmit power will be used for this particular beam. 
%it is not hard to see that the ET employs a pairwise antenna activation policy in each slot $ k $, that is, it only activates a pair of antennas for the generation of this beam.
%Consider that the first antenna and the $k$-th antenna, where $k \in \brparen{2,\ldots,K}$, are activated when $ l \neq L+1 $. 

Next, we will provide further insights on the design of the codebooks by looking into the estimation process.
To this end, by using \eqref{power}, the RSSI at the $ i$-th ER for the $ l$-th ($ \leq L $) element (beam) of $\mathrm {\bf {B}}_v $ can be written as
\begin{alignat} {2}
\mathrm R_{i,l}^v
&=\alpha_{i,v} + \beta_{i,v} \cos\paren{\theta_l+\phi_{i,v}} + z_i,
\label{eq:ch rssi_2} 
\end{alignat} 
where 
$ \alpha_{i,v} = \frac{ P}{4}(|h_{i,1}|^{2} + |h_{i,v}|^{2}), $
$ \beta_{i,v} = \frac{ P}{2}|h_{i,1}||h_{i,v}|, $
and $\phi_{i,v}= \delta_{i,v} - \delta_{i,1}.$ Although we have assumed a quasi-static block-fading channel, due to the effect of noise, the RSSI value will change from one measurement to the other. We have used random variable $ z_i $ to represent this effect. More specifically, $ z_i $ captures the effect of all noise related to the measurement process such as noise in the channel, circuit, antenna matching network and rectifier, and we assume the random variables to be i.i.d. additive Gaussian having zero mean and variance $ \sigma^2 $. Therefore, we assume that in a given transmission block, the randomness in \eqref{eq:ch rssi_2} is caused only by $ z_i $.

It can be seen from \eqref{eq:ch rssi_2} that $  \mathrm R_{i,l}^v $ depends on three unknown parameters $ \alpha_{i,v}$, $\beta_{i,v}$, and $ \phi_{i,v}$. Thus, the parameter vector for the estimation process can be written as $\vec \varphi = [\alpha_{i,v} \enspace \beta_{i,v} \enspace \phi_{i,v}]^{\top} $. For a given $v \in \brparen{2,\ldots,K}$, the ET will receive $L$ feedback values in the form of \eqref{eq:ch rssi_2} from each ER, and these feedback values will be utilized to estimate $\phi_{i,v}$ for each $i \in \brparen{1,\ldots,N}$. This means, the pairwise antenna activation is used to estimate the phase information, and these estimates give us enough information to perform EGT. However, 
to perform optimal beamforming, we need amplitude information as well. We use the $L+1$-th beamforming vector in each codebook for this purpose, and the amplitude information can be obtained by estimating $ \alpha_{i,v}$ and/or $\beta_{i,v}$. As shown later, estimating either $ \alpha_{i,v}$ or $\beta_{i,v}$ is sufficient for our requirement, thus, we will estimate $ \alpha_{i,v}$ without any loss of generality. 

It should be highlighted that there is a reason behind selecting $ \theta_l =  \frac{2(l-1)\pi}{L} $ for $l \in \brparen{1, \ldots, L}$ as well. We have selected $\brparen{\theta_l}_{l=1}^L$ values in our codebooks in a manner such that the estimators of all three parameters of interest achieve the Cramer-Rao lower bound (CRLB). The CRLB is the best performance that an unbiased estimator can achieve as it gives a lower bound on the variance of an unbiased estimator. The analysis in \cite{our_tsp} shows this rigorously for the phase estimates (which are the estimates of interest in \cite{our_tsp}), and by using a similar approach, we can show that the same selection does the best in estimating the other two parameters of the parameter vector as well. Also, it can be shown by using the Fisher information matrix of $\vec \varphi$, that $L \geq 3$ for the estimation process to be possible \cite{our_tsp}. We skip the proof details to avoid repetition of similar results. 

Now, let us focus on the estimation of parameters. Based on the assumption that the effective  noise is i.i.d. Gaussian, estimating $\phi_{i,v}$ and $ \alpha_{i,v}$ for a given $i$ and $v$ becomes a classical parameter estimation problem. A maximum likelihood estimate of these parameters can be obtained by finding the values of $\phi_{i,v}$ and $ \alpha_{i,v}$ that minimize 
\begin{eqnarray}
\mathrm E \defeq \sum_{l=1}^{L} \sqparen{\mathrm R_{i,l}^v - \paren{\alpha_{i,v} + \beta_{i,v} \cos\paren{\theta_l+\phi_{i,v}}} }^{2}. \label{ml}
\end{eqnarray}  
These ideas are formally presented through the following theorem. 
\begin{theorem} \label{Thm:esti1}
For all $i \in \brparen{1, \ldots, N}$ and $v \in \brparen{2, \ldots, K}$, the estimates of $ \phi_{i,v} $  and $ \alpha_{i,v} $ are given by  
\begin{gather}
\hat	 \phi_{i,v} = \tan^{-1}\paren{\frac{\displaystyle  -\sum_{l=1}^{L} \mathrm R_{i,l}^v \sin{\paren{\theta_l}} }           
	{\displaystyle \sum_{l=1}^{L} \mathrm R_{i,l}^v \cos{\paren{\theta_l}}}	
},
\label{estimation_phi} 
\end{gather}
and 
\begin{eqnarray}
\hat{\alpha}_{i,v} = \displaystyle  \sum_{l=1}^{L} \frac{\mathrm R_{i,l}^v}{L}, \label{estimation_alpha}
\end{eqnarray}
%\begin{equation}
%\hat{\alpha}_{i,v} = \displaystyle  \sum_{l=1}^{L} \frac{\mathrm R_{i,l}^v}{L}, \label{estimation_alpha}
%\end{equation}
%\begin{gather}
%% \hat{\alpha}_{i,v} = \frac{\displaystyle  \sum_{l=1}^{L} \mathrm R_{i,l}^v  }           
%% 	{\displaystyle L}	,
% \hat{\alpha}_{i,v} = \displaystyle  \sum_{l=1}^{L} \mathrm R_{i,l}^v / L,      	
%\label{estimation_mag}
%\end{gather}
respectively, where $ \theta_l =  \frac{2(l-1)\pi}{L} $.
\end{theorem}
\begin{IEEEproof}
Differentiating $\mathrm E$ in \eqref{ml} with respect to  $\phi_{i,v}$, and setting it equal to zero gives us
\begin{multline}
	\sum_{i=l}^{L} \mathrm{ R}_{i,l}^v\sin{(\theta_{l}+\phi_{i,v})} = \alpha_{i,v} \sum_{l=1}^{L} \sin{(\theta_{l}+\phi_{i,v})} \\ + \frac{\beta_{i,v}}{2} \sum_{l=1}^{L} \sin{[2(\theta_{l}+\phi_{i,v})]}.
	\label{eq:estimation_phi}
\end{multline}
It can be seen that to estimate $\phi_{i,v}$, we need estimates of $\alpha_{i,v}$ and $\beta_{i,v}$. However, due to the definition of $\theta_l$ through the CRLB analysis, we get $$\sum_{l=1}^{L} \sin(\theta_{l}+\phi_{i,v})=\sum_{l=1}^{L} \sin{[2(\theta_{l}+\phi_{i,v})]}=0, $$ using series of trigonometric functions \cite{ryzhik}. Therefore, \eqref{eq:estimation_phi} simplifies into $$ \sum_{l=1}^{L} \mathrm{ R}_{i,l}^v\sin{(\theta_{l}+\phi_{i,v})} = 0. $$ Expanding $\sin{(\theta_{l}+\phi_{i,v})}$ allows us to obtain \eqref{estimation_phi}.
Differentiating $\mathrm E$ in \eqref{ml} with respect to  $\alpha_{i,v}$, and setting it equal to zero gives us
\begin{equation*}
	\sum_{i=l}^{L} \mathrm{ R}_{i,l}^v - L\alpha_{i,v}  - \frac{\beta_{i,v}}{2} \sum_{l=1}^{L} \cos\paren{\theta_l+\phi_{i,v}}=0.
\end{equation*}
We obtain \eqref{estimation_alpha} since $\sum_{l=1}^{L} \cos\paren{\theta_l+\phi_{i,v}}=0$, which completes the proof.
\end{IEEEproof}

Note that the results in Theorem \ref{Thm:esti1} are  simple,  easy to calculate,
and require minimal processing. We should highlight that as shown in the proof, the manner in which we  selected $\brparen{\theta_l}_{l=1}^L$ have indirectly led to the simplifications of these results. Ambiguity resolution in $\hat	 \phi_{i,v}$ can be done using similar techniques discussed in \cite{our_tsp}.

Next, let us focus on the $(L+1)$-th vector of each codebook. We have $\alpha_{i,v} = \frac{ P}{4}(|h_{i,1}|^{2} + |h_{i,v}|^{2})$, which we have already estimated. For optimal beamforming, we need $|h_{i,v}|$, and to extract this from $\alpha_{i,v}$, we need to know $|h_{i,1}|$. When $ l =L+1 $, we transmit using the first antenna only, and the corresponding RSSI value is given by %\vspace{-0.2cm}
\begin{alignat} {2}
\mathrm R_{i,(L+1)}^v =  \frac{ P}{2} |h_{i,1}|^{2} + z_i.
\label{eq:ch rssi_3} 
\end{alignat}
Estimating $|h_{i,1}|$ can be done using the same concepts as earlier, and these estimates will allow us to recover estimates of $\brparen{|h_{i,v}|}_{v=2}^K$. We present the results through the following theorem and we skip the proof for brevity. 
\begin{theorem} \label{Thm:esti2}
For all $i \in \brparen{1, \ldots, N}$ and $v \in \brparen{2, \ldots, K}$, the estimates of $ |h_{i,v}| $ are given by  
\begin{equation}
|\hat{h}_{i,v}| = \sqrt{ \frac{4}{P} \hat \alpha_{i,v} - |\hat h_{i,1}|^{2} },
 \label{estimation_h}
\end{equation}
where $$ |\hat h_{i,1}|^{2} = \frac{2}{P(K-1)} \sum_{v=2}^{K}  \mathrm R_{i,(L+1)}^v.  $$
\end{theorem}

Now we have sufficient information to perform WET using optimal beamforming, and also to cluster the ERs  using the Lloyd's Algorithm. We should note that if $K=2$, the ET will only receive one feedback value of the form in \eqref{eq:ch rssi_3}. For this case, we will have to repeat the $(L+1)$-th beamforming vector to get some more feedback values to facilitate the estimation process of $|h_{i,1}|$. Next, we  focus on solving the optimization problem to select a beamforming vector for the WPB stage.   %\vspace{0.3cm}

%For a given $k$, we can do this for all $i \in \brparen{1,\ldots,N}$, and repeating this for all $k \in \brparen{2, \ldots, K}$ gives us $\brparen{|h_{i,k}|}_{k=1}^K$ and  $\brparen{\phi_{i,k}}_{k=2}^K$.  

%\section{Clustering of ERs using vector quantization} \label{clu}
%
%
%In this section, we will use channel estimates $\brparen{ \hat {\vec h}_i }_{i=1}^{N}$ to define clusters of ERs through vector quantization. More precisely, we will assign $ N $ ERs into $ M (< N)$ clusters, so that it minimizes the sum of euclidean distances between cluster centroids $\brparen{  {\vec h}_m }_{m=1}^{M}$ and channel estimates, \textit{i.e.}, 
%\begin{equation*}
%\argmin_{\mathcal{C}_m}  \sum_{m=1}^{M} \sum_{\hat {\vec h}_i \in \mathcal{C}_m} \|\hat {\vec h}_i - \vec h_m^c \|^2_{\mathrm F},
%\end{equation*}
%where $ \mathcal{C}_m $ is the set of channel estimates that belong to cluster $  m \in \{1,\dots,M \} $. In this paper, Lloyd's algorithm \cite{lloyd} is adopted to solve this problem, and it can be summarized as follows. 
%\begin{equation*}
%\mathcal{C}_m = \{i: \|\hat {\vec h}_i - \vec h_m^c \|^2_{\mathrm F} \enspace \leq \|\hat {\vec h}_i - \vec h_l \|^2_{\mathrm F}, l\neq m, i=1,\dots ,N\}
%\end{equation*}

%However, for a random data set, 

%For K-means clustering algorithm, choosing the number of clusters $ M $ is crucial to the performance of clustering. 

\section{Solution to the optimization problem} \label{opt}

The problem in \eqref{Optimization Problem-MU} is convex, but it is complex due to $ \mathrm C1 $ having infinitely many inequalities.
Therefore, this problem can be alleviated by transforming $ \mathrm C1 $ into a linear matrix inequality (LMI) \cite{MIQ}, 
%\cite{MIQ}
and this is possible by applying the S-procedure \cite{SPro}. These ideas are formally presented through the following lemma. 
\begin{lemma} \label{Lemma:LMI}
	The equivalent LMI of constraint $ \mathrm C1 $ in (\ref{Optimization Problem-MU}) is given by 
%	$ \mathrm \mathrm {\bf T}_{i}(\mathrm {\bf C}_{\vec {xx}},t, \mu_i) \succeq 0 \ \ \forall i \in \mathcal{Q}^\star $,
	\begin{equation}
	\begin{aligned}
	\mathrm \mathrm {\bf T}_{i}(\mathrm {\bf C}_{\vec {xx}},t, \mu_i) \succeq 0 \ \ \forall i \in {S}^\star, \nonumber 
	\end{aligned}
	\end{equation} 	
	where
	\begin{equation*}
	\mathrm {\bf T}_{i}(\mathrm {\bf C}_{\vec {xx}},t, \mu_i) %\hspace{-1mm}
	= \\ %\hspace{-2.2mm}
	\begin{bmatrix}
	\mu_i  \mathrm {{{\bf{I}}}}_K+  \mathrm {\bf C}_{\vec {xx}}    &  \mathrm {\bf C}_{\vec {xx}} \vec h_i\\
	\vec h_i^{\dag} \mathrm {\bf C}_{\vec {xx}} &  \vec h_i^{\dag} \mathrm {\bf C}_{\vec {xx}}\vec h_i-t-\mu_i \varepsilon_i^2 \\
	\end{bmatrix} ,
	\end{equation*}
	and $ \mu_i \geq 0$ is a real-valued variable.
	\label{LMI}
\end{lemma}
\begin{IEEEproof}
	For $ g=1,2 $, let $ f_g(\vec \eta_i) $ be defined as 
	\begin{alignat} {2}
	f_g(\vec \eta_i)
	&= \vec \eta_i^{\dag} \mathrm{ \bf{A}}_g \vec \eta_i  + 2\text{Re} \{ \mathrm{\bf b}_g^{\dag} \vec \eta_i \} + c_g, \nonumber
	\end{alignat}
	where $\mathrm{ \bf{A}}_g \in \mathbb{C}^{K\times K}  $, $\mathrm{ \bf{b}}_g \in \mathbb{C}^{K\times 1}  $, and $ c_g \in \mathbb{R} $. According to \cite{SPro}, the deduction (implication) $ f_1(\vec \eta_i) \leq 0 \Rightarrow f_2(\vec \eta_i) \leq 0 $ holds if and only if there exists a $ \mu_i \geq 0$ such that 
	\begin{equation*}
	\begin{bmatrix}
	\mathrm{\bf A}_2   &  \mathrm{\bf b}_2\\
	\mathrm{\bf b}_2^{\dag} &  c_2 \\
	\end{bmatrix} 
	\preceq  \mu_i\\
	\begin{bmatrix}
	\mathrm{\bf A}_1   &  \mathrm{\bf b}_1\\
	\mathrm{\bf b}_1^{\dag} &  c_1 \\
	\end{bmatrix},
	\end{equation*}
	provided there exists a point $ \vec {\hat \eta_i} $ such that $ f_1(\vec {\hat \eta_i}) < 0 $. Now, with the focus of applying the S-procedure, we write $ \mathrm {C1} $ as the
	following implication:
\begin{multline} 
{\vec \eta_i^{\dag} \mathrm {{{\bf{I}}}}_K \vec \eta_i }  \leq \varepsilon_i^2
\Rightarrow  -\vec \eta_i^{\dag} \mathrm {\bf C}_{\vec {xx}} \vec \eta_i  - \\ 2\text{Re} \{ \vec h_i^{\dag} \mathrm {\bf C}_{\vec {xx}} \vec h_i \vec \eta_i \}  - \vec h_i^{\dag} \mathrm {\bf C}_{\vec {xx}} \vec h_i + t\leq 0. \label{impli}
\end{multline}
	Using the definition of S-procedure, writing \eqref{impli}  as
	\begin{equation*}
	\begin{bmatrix}
	-\mathrm {\bf C}_{\vec {xx}}   &  -\mathrm {\bf C}_{\vec {xx}} \vec h_i\\
	-\vec h_i^{\dag} \mathrm {\bf C}_{\vec {xx}} &  -\vec h_i^{\dag} \mathrm {\bf C}_{\vec {xx}} \vec h_i+ t \\
	\end{bmatrix} 
	\preceq  \mu_i\\
	\begin{bmatrix}
	\mathrm {{{\bf{I}}}}_K  &  0\\
	0 &  -\varepsilon_i^2 \\
	\end{bmatrix}
	\end{equation*}
	\begin{equation*}
	\Rightarrow 0
	\preceq  \\
	\begin{bmatrix}
	\mu_i  \mathrm {{{\bf{I}}}}_K+  \mathrm {\bf C}_{\vec {xx}}    &  \mathrm {\bf C}_{\vec {xx}} \vec h_i\\
	\vec h_i^{\dag} \mathrm {\bf C}_{\vec {xx}} &  \vec h_i^{\dag} \mathrm {\bf C}_{\vec {xx}}\vec h_i-t-\mu_i \varepsilon_i^2 \\
	\end{bmatrix}
	\end{equation*}
	completes the proof.
\end{IEEEproof}

\begin{figure}[t] 
	\centering {\includegraphics[trim = 12mm 5mm 0mm 8mm, clip, scale=0.58]{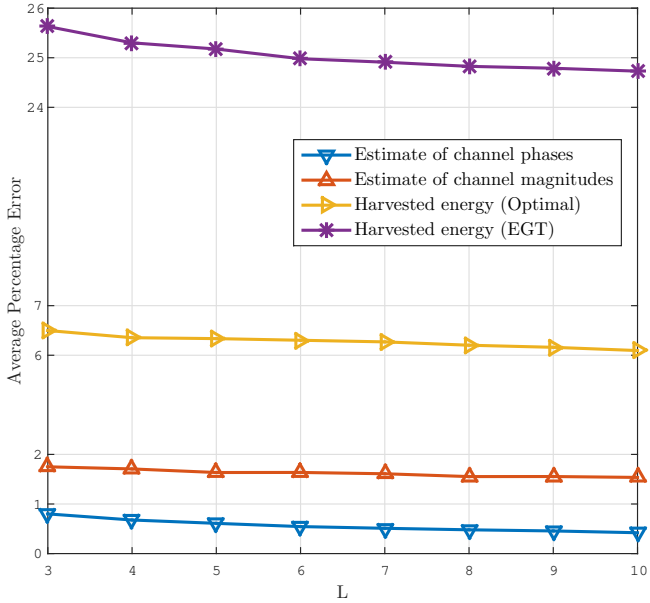}} 
	\caption{The behavior of the error percentage, for $K=4$ and SNR $ = $ 20dB.}	 
	\label{csi} 
\end{figure}

%\begin{figure}[t]
%	{\includegraphics[trim = -65mm 0mm 0mm 10mm, clip, scale=0.45]{mcrlb2}} 
%	\caption{The behavior of the $ \mathrm{MCRLB}_{\phi}$ with N when $\beta=\sigma=1$.}	 
%	\label{mcr}
%\end{figure}

Using this lemma, we can write the following equivalent optimization problem. 
\begin{equation}
\begin{aligned}
& \underset{\footnotesize \mathrm {\bf C}_{\vec {xx}} \succeq 0, t, \mu_i  }{\text{maximize}}
& & t  \\
& \text{subject to}
& & \mathrm C1:  \enspace \mathrm {\bf T}_{i}(\mathrm {\bf C}_{\vec {xx}},t, \mu_i) \succeq 0 \ \ \forall i \in {S}^\star\\
& &&  \mathrm C2: \mathrm{tr}(\mathrm {\bf C}_{\vec {xx}}) \leq P. 
\end{aligned} \label{sdp}
\end{equation}
This is a  semidefinite programming (SDP) problem and it can be easily solved by
using numerical convex program solvers such as CVX \cite{cvx}, and we have already established that optimality is achieved when $\mathrm{rank}(\mathrm {\bf C}_{\vec {xx}})=1$ \cite{rank_one}.

\section{Simulation Results and discussion} \label{results}

In this section, we present two sets of simulation results/numerical evaluations to highlight the two main contributions of this paper. In both simulations, the random channel amplitudes are assumed to be uniform between $0.1$ and $1$,  and averaging is done over $1000$ iterations. Firstly, we focus on the channel estimation. To this end, Fig. \ref{csi} illustrates how the  average error percentage changes with the amount of feedback, focusing on one ER. We can see that the phase estimation error and the magnitude estimation error are both  very low. The two graphs on harvested energy represent the average loss in harvested energy due to opting for RSSI based channel estimation, instead of optimal beamforming with perfect CSI. We can see that the loss is rather acceptable given the practicality of the proposed method compared to having perfect CSI at the ET. We can also see that there is a significant improvement of going for optimal beamforming using the channel estimation techniques in this paper, compared to the EGT beamforming used in \cite{our_rssi}. For the selected parameters in this simulation, the improvement is approximately $20\%$. Also, Fig. \ref{csi} illustrates that larger $ L $ values yield a higher channel estimation precision. However, larger $ L $ will increase the time spent in training, which will eventually reduce the time for WPB. This may lead to a reduction in the total transferred energy. Therefore, the selection of $ L $ affects the system performance greatly. We leave this for future work. 

\begin{figure}[t] 
	\centering {\includegraphics[trim = 12mm 5mm 0mm 8mm, clip, scale=0.56]{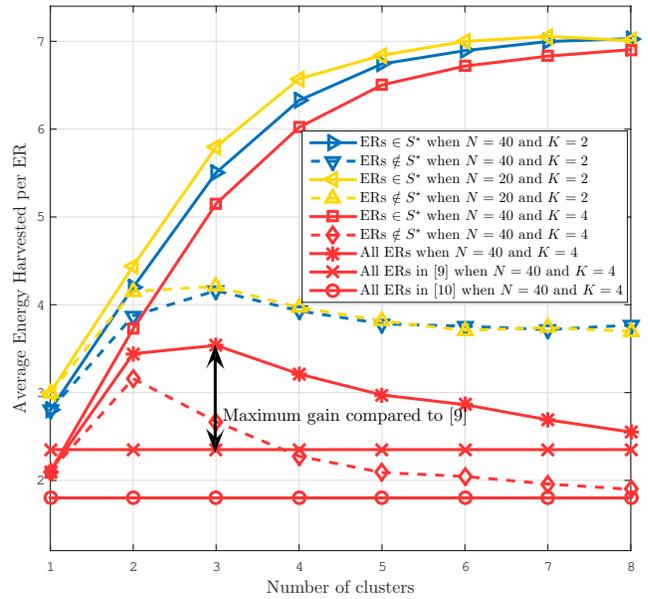}} %0.45
	\caption{The behaviour of the harvested energy.}	 
	\label{opp} 
\end{figure}

Fig. \ref{opp} illustrates the effect of the number of clusters on the average energy harvested per ER. Note that the number of clusters being equal to one is equivalent to having no clustering, $\ie$, we try to maximize the minimum harvested energy among all $ N $ ERs. It is not hard to see that clustering is certainly useful.  For example, when $Q=3$, $N=40$ and $K=4$, we get an approximately $75\%$ improvement in the average energy harvested per ER due to clustering, with the selected parameters. It is rather obvious that ERs in ${S}^\star$ should harvest more energy, but for a given $Q$, the energy harvested by the ERs has decreased with both $N$ (due to having a lesser number of ERs in the selected cluster percentage wise) and $K$ (due to the beam being more directive). It is also interesting to note that the proposed opportunistic scheduling policy outperforms both \cite{oppor1} and \cite{oppor2} for the selected parameters. It is rather intuitive that the proposed scheme will achieve fairness. Therefore, we omit such simulation results.

\section{Conclusions} \label{conclusions}

This paper has proposed a novel channel estimation method, and an opportunistic scheduling policy to be used in
a WET system consisting of multiple low complex ERs. In the training stage, the ET transmits using a set of predefined codebooks, and each ER feeds back corresponding RSSI values to the ET. These values are used for channel estimation. Based on the channel phase 
estimates, the ERs are grouped into clusters, and the most dense cluster is selected for dedicated  WET. 
The beamformer that maximizes
the minimum harvested energy among all ERs in the selected
cluster is found by solving a convex optimization problem. This beamformer is used to transfer power to the ERs using optimal beamforming, while achieving fairness over time.  Insightful simulation results and numerical evaluations have been presented to
validate  the performance gains that can be achieved from the
proposed schemes.

\balance

\footnotesize {\bibliography{bibfile}}

\end{document}